\documentclass[journal]{IEEEtran}
\usepackage[usenames]{color}
\usepackage{epsfig}
\usepackage{graphics}
\usepackage{amsmath}
\usepackage{amssymb}
\usepackage{multirow}
\usepackage{cite}
\usepackage{array}
\usepackage{pslatex} 
\usepackage{subcaption}
\usepackage{caption}
\usepackage{multicol}
\usepackage{lipsum}
\usepackage{graphicx}  
\usepackage{setspace}
\usepackage{tikz}
\usepackage{hhline}
\usepackage{mathtools}
\usepackage[letterpaper]{geometry}
\definecolor{mypink1}{rgb}{0.0, 0.0, 1.0}
\definecolor{red1}{rgb}{1.0, 0.01, 0.24}
\definecolor{green1}{rgb}{0.0, 0.5, 0.0}
\usepackage{titlesec}

\titlespacing*{\section}
{0pt}{1ex plus 0.2ex minus .1ex}{1ex plus .1ex}
\titlespacing*{\subsection}
{0pt}{1ex plus 0.2ex minus .1ex}{1ex plus .1ex}

\ifCLASSINFOpdf

\else
\fi
\begin{document}
	%
	\title{Re-configurable Intelligent Surface-based VLC Receivers Using Tunable Liquid-crystals: The Concept}
	%
	%
	%
	
	\author{Alain R. Ndjiongue, Telex M. N. Ngatched, Octavia A. Dobre, and Harald Haas 
	\thanks{A. R. Ndjiongue, T. M. N. Ngatched, and O. A. Dobre are with the Faculty of Engineering and Applied Science, Memorial University of Newfoundland, Canada. \\ Harald Haas is with the LiFi Research and Development Center, Dpt. Electronic and Electrical Engineering, the University of Strathclyde, Glasgow, U.K.}}
	\maketitle
	
	\begin{abstract}
Visible light communication (VLC) enables access to huge unlicensed bandwidth, a higher security level, and no radio frequency interference. With these advantages, VLC emerges as a complementary solution to radio frequency communications. VLC systems have primarily been designed for indoor scenarios with typical transmission distances between 2 and 5 m. Different designs would be required for larger distances. This paper proposes for the first time the use of a liquid crystal (LC)-based re-configurable intelligent surface (RIS) on improving the VLC signal detection and transmission range. An LC-based RIS presents multiple advantages, including the tunability of its photo-refractive parameters. Another advantage is its light amplification capabilities when under the influence of an externally applied field. In this paper, we analyze an LC-based RIS structure to amplify the detected light and improve the VLC signal detection and transmission range. Results show that mixing LC with 4 to 8 wt\% concentration of a dye such as the terthiophene (3T-2MB) improves the VLC transmission range of about 0.20 to 1.08 m. This improvement can reach 6.56 m if we combine 8 wt\% concentration of 3T-2MB and 0.1 wt\% concentration of trinitrofluorenone.
	\end{abstract}
	
	\begin{IEEEkeywords}
Visible light communication (VLC), re-configurable intelligent surface (RIS), dye-doped and polymer liquid crystals, terthiophene (3T-2MB), trinitrofluorenone (TNF), VLC transmission range expansion, radiative transfer equation (RTE), Beer-Lambert law.
	\end{IEEEkeywords}
	%
	\IEEEpeerreviewmaketitle     
	\section{Introduction}
One of the main impediments of visible light communication (VLC) systems is its short transmission range imposed by the source lighting area. Using a convex lens in VLC receivers to focus the incident light on the photodetector (PD) surface worsens this impediment as it creates losses. Traditional VLC receivers use convex, spherical, or compound parabolic concentrators to steer the detected light. But, due to etendue, field-of-view is traded-off against higher received power. These concentrators are made of the same material and have different shapes, and most employ a glass slab or other transparent materials characterized by high transparency, low or in-existent scattering and absorbance, to allow light to pass through with a constant intensity. These convex lenses have a transmittance of about 90\% for waves in the visible spectrum. For example, at 0$^o$ incidence of the incoming light rays, most glasses with a refractive index around 1.5 have a decrease of about 2 to 4\% and 5 to 10\% of the incident light power for clear and prismatic glasses, respectively. These values can reach 30\% for heat-absorbing glasses and represent an intensity loss due to reflection at the lens's upper surface. Their impacts become significant at the receiving edge, where the detected signal is already considerably attenuated. This paper deals with this problem using liquid crystal (LC)-based re-configurable intelligent surface (RIS) and introduces solutions to improve the VLC transmission range.
	
Due to the tunability of their physico-chemical characteristics, the RIS technology is attracting significant research interests. For example, in radio frequency, the reflection properties of RIS is exploited to solve the dead zone problem and enhance signal detection. In VLC, the refractive attributes of RIS can be used to steer the incoming light beam and mitigate intensity loss. The latter represents the focus of this paper.
	
LCs are good candidates for the optical RIS technology because of the tunability of their refractive characteristics, including emission and attenuation coefficients, and the refraction index. This is due to the externally applied field, which creates a variation of these parameters through population change in a doped substance. As a result, incident light steering and amplification can be obtained. Besides displays and other previous LC applications, including scanning, safety light amplification for industrial use, and measurement, the semiconductor packaging industry exploits the LC polymer to solve packaged device degradation, which occurs when water or other substances accumulate in an open area \cite{5944934}. A recent development also highlights that the next generation automobile considers LC as the heart of new light detection, ranging (LIDAR), and remote sensing technologies. Recently developed self-driving cars use LC-based LIDARs \cite{rablau2019lidar}. In this paper, we utilize LC to design a RIS-based VLC receiver with low-light amplification capabilities, with the goal of enhancing signal detection and improving the transmission range.
	\begin{figure*}
		\centering
		\includegraphics[width=0.73\textwidth]{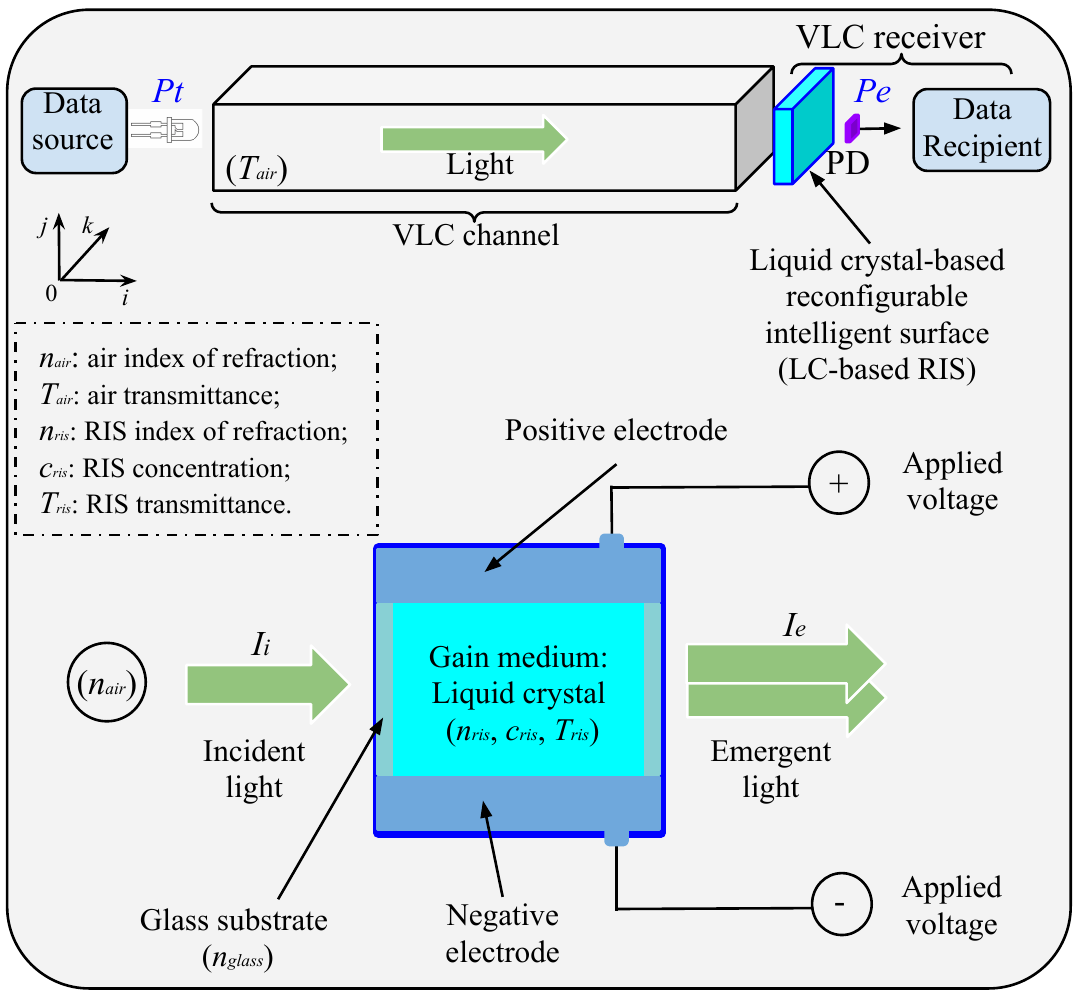}
		\caption{A model of VLC systems using an LC-based RIS in the receivers. Top: VLC transmission systems using RIS-based receiver. Bottom: Illustration of light amplification through the RIS element.}
		\label{fig:model}
	\end{figure*}
	
Two categories of phenomena affect the incident light when it travels through the air or any material. The direction-related phenomena, which include scattering, reflection, refraction, and intensity-related phenomena occurring on the signal power, such as attenuation and emission. Absorption and scattering contribute to signal attenuation in the considered media, and are described by the Beer-Lambert law, while emission refers to the part of light passing through the sample. When the emission coefficient is null, solutions to the radiative transfer equation (RTE\footnote{RTE is a mathematical description of absorption, emission, and scattering phenomena, which affect light propagation through a sample.}) lead to Beer-Lambert law, which focuses on the light attenuation aspect. The high values of this parameter may result in total light extinction. On the other hand, light amplification concerns both attenuation and emission parameters. LC-based RIS can be used in a VLC receiver to prevent losses due to convex lenses and provide light amplification to enhance signal reception. 
	
One of the most used low-light amplifiers exploits silicon photomultipliers, a combination of PDs, resistors, and operational amplifiers. It first converts the light into an electric current, then proceeds to its amplification. Even though the amplification gain is tunable, this type of system does not allow direct light re-transmission. In addition to its tuning smoothness, an LC-based RIS low-light amplifier does pure light amplification and presents a twofold-impact on the incident beam, namely, steering and amplification \cite{alain}. Employing this type of amplifier in VLC receivers has not yet been proposed or explored in the open literature. Hence, in this paper, we discuss selected LC mixtures to create a RIS-based low-light amplifier and enhance VLC signal detection. We tune the mixture properties to reduce its attenuation coefficient and increase the emission parameters through an externally applied electric field. This paper's main goal is to show that the use of LC-based RIS in a VLC receiver has the potential to unquestionably enhance its signal detection capabilities. To this end, we make the following contributions: (\textit{i}) we revisit the RTE, Einstein coefficients, and Beer-Lambert law, and highlight their applications; (\textit{ii}) we propose the principle of light amplification based on tunable RIS attenuation coefficient, and analyze it to provide the different approaches to use LC-based RIS samples in VLC systems; (\textit{iii}) finally, we suggest future research directions.
	\section{RIS-based VLC System Model, RTE Basic Concept, and Principles}
	\subsection{A VLC System with LC-based RIS Structure}
Figure~\ref{fig:model} depicts a model and principle of a VLC system that incorporates an LC-based RIS element in the receiver. Its top-part shows a data and light source, the channel, an LC-based RIS element, and the PD. A light-emitting diode, acting as the light source, converts the data into light. The resulting signal lights up the environment and carries over the information by propagation through the transmission medium. In the receiver, we replace the traditional concentrator, which is based on convex lens, with an LC-based RIS structure placed before the PD. Its role is to perform incoming light conditioning before it reaches the PD. Finally, the PD converts detected lights into electrical signals for further processing. The bottom part of the figure illustrates the principle of low-light amplification using an LC-based RIS. It shows that light intensity emerging from the RIS sample, $ I_e $, is greater than the incident light intensity, $ I_i $. The LC-based RIS transmittance, $ T_{ris} $, characterizes the RIS element, while $ T_{air} $ describes the air-based part of the channel. The overall system transmittance, which corresponds to the VLC channel DC gain, is the product of $ T_{air}$ and $T_{ris} $. Tuning the VLC transmittance implies controlling the channel DC gain. However, it is not straightforward to control the air-based channel characteristics. Hence, this paper focuses on tuning the LC-based RIS properties to solve the attenuation in the air-based part of the channel and control the light intensity falling on the PD's surface. $ T_{ris} $ typically depends on the LC physico-chemical characteristics, which are parts of the RTE and describe the change in light intensity as the radiation propagates, depending on the light direction, frequency, absorption, and emission phenomena.
	\subsection{The RTE and Beer-Lamber Law: Concept of Gain Medium}
Heating atoms in matter leads to a generation of little packets of energy, which constitute light. During propagation, this light transfers its energy from one point to another through emission, absorption, and scattering phenomena. These can take place over four radiation parameters: energy-flux, direction, frequency, and polarization, and are likely to influence the course of the signal, resulting in intensity and direction change. However, propagation occurs with a multi-order uniform spreading depending on the nature of the medium. When the generated photons reach a different medium, their energies affect the atoms in the medium. As a result, the medium particles can fully absorb the photons (optically opaque materials) or allow them through (optically transparent materials). The absorption coefficient, $\alpha_{\lambda}$, and emission coefficient, $\epsilon_{\lambda}$, respectively define these two scenarios and represent the Einstein coefficients of the RTE \cite{bernath2020spectra}. The RTE defines a relationship between these parameters, the light intensity, and the sample depth. These coefficients are determined using the Planck constant, wave frequency, atom density in lower and upper energy states for the atomic spectra, combined with Einstein coefficients for photon absorption, spontaneous and induced emissions \cite{bernath2020spectra}. Thus, the intensity of a light traveling through an LC-based RIS structure obeys to the RTE, in which the Kirchhoff law defines the source function given by the Planck's equation. Consequently, the LC-based RIS element is entirely opaque if $\epsilon_{\lambda} = 0$, which leads to a total light extinction, or perfectly transparent if $\alpha_{\lambda} = 0$. Absorption and scattering phenomena lead to light attenuation, which is an exponential suppression governed by the Beer-Lambert law, where the attenuation coefficient is defined as the sum of absorption and scattering coefficients. Generally, attenuation and emission coefficients are intrinsic characteristics of the material used to build the sample. Their combination always leads to a total attenuation of light after a given distance. Depending on the material's atomic constitution, it may absorb the photon and allow a re-emission to happen. This process occurs when the incoming photon excites the sample's atom and forces it to change its excitation level. These two operations result in absorption or spontaneous emission depending on the polarity of charges in the sample.
	\subsection{Principle of Low-light Amplification Using LC-based RIS}
In most light amplifiers, an atomic population change is created through pumping, which results in the inversion of the absorption coefficient sign. This mechanism, also called stimulated emission, has revolutionized the laser industry during the last three decades, with many applications including cutting, graving, distance measurement, and medicine, to mention only a few. In LC-based RIS VLC receivers, we exploit this principle to amplify a low-intensity detected incident light in order to strengthen the VLC signal for better detection or re-transmission.  
	\begin{figure}
		\centering
		\includegraphics[width=0.44\textwidth]{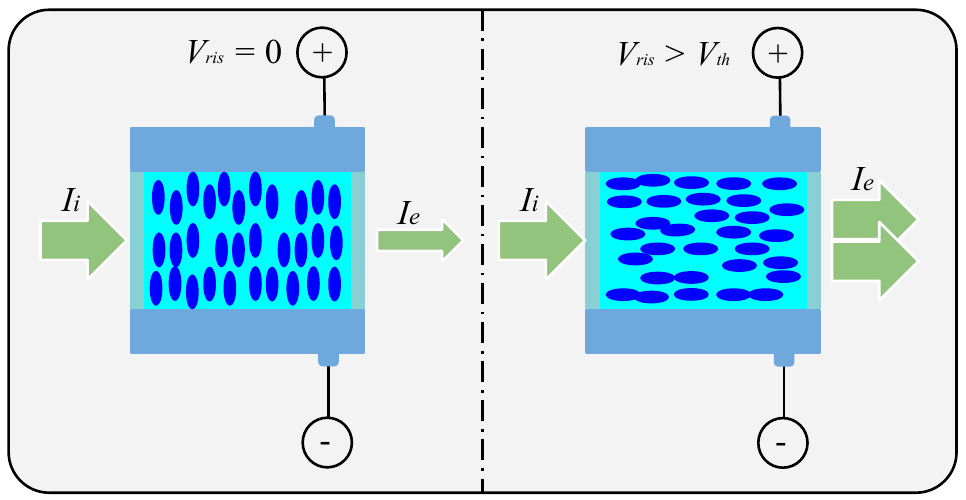}
		\caption{Low-light amplification mechanism using an LC-based RIS sample.}
		\label{fig:Ampli_LC}
	\end{figure}
	
Applying a positive or negative variable external electric or magnetic fields to the LC cell leads to the tuning of its photo-refractive parameters, and has undeniable effects on the incident light beam \cite{hsu2019electric, adamow2020light}. In most cases, the LC structure amplifies the incident light. An external electric field applied to an LC cell using perylene-based dye results in an amplified emerged light with a considerable gain. In dynamic image amplification, ferroelectric LC containing photoconductive compound chiral may exhibit a gain coefficient, $\Gamma_{\lambda}$, exceeding 1200 cm$^{-1}$. Some biosensors employ LC to detect bio-molecules by amplifying the signal emitted by these \cite{luan2020application}. Although the applied electrical field influences the light intensity, these LC-based structures amplify light at an optimal LC cavity depth for maximal signal amplification. For example, with a depth, $ x $, of 4.7 $\mu$m, a manually-fabricated sample may provide an emerged light with a significant gain \cite{hsu2019electric}.
	\section{Low-light Amplifier Using an LC-based RIS}  
After reflecting on the upper surface of the LC-based RIS, the remaining light undergoes the three phenomena described by the general solution to the RTE, which lead to a loss, gain, and redistribution of energy, due to absorption, emission, and scattering processes, respectively. These represent an explicit expression and characteristics of the material in LC-based RIS cells. The absorption coefficient, $\alpha_{\lambda}$, in combination with the distance over which the light travels, defines the amount of light absorbed by the structure. It would be ideal to find an LC substance with lower absorption, scattering, and higher emission coefficients. However, this is not common. Thus, the best solution is to induce a population change in the material so that the absorption process becomes an emission. This can be readily achieved with the help of an externally applied electric field. Figure~\ref{fig:Ampli_LC} illustrates this amplification operation with a two-state scenario, whereby the left-hand side shows attenuation of incident light when there is no external field. In this scenario, the LC molecules align themselves in parallel with the glass surface. On the right-hand side, the figure depicts the amplification scenario when an externally applied voltage, $ V_{ris} $, is greater than the switching potential, $ V_{th} $ \cite{7870642}. Here, the LC molecules change direction and turn vertically to the glass. Biased by the externally applied field, the LC material generates photons identical to the incident ones and moving into the same direction.
		
\begin{figure}
		\centering
		\includegraphics[width=0.45\textwidth]{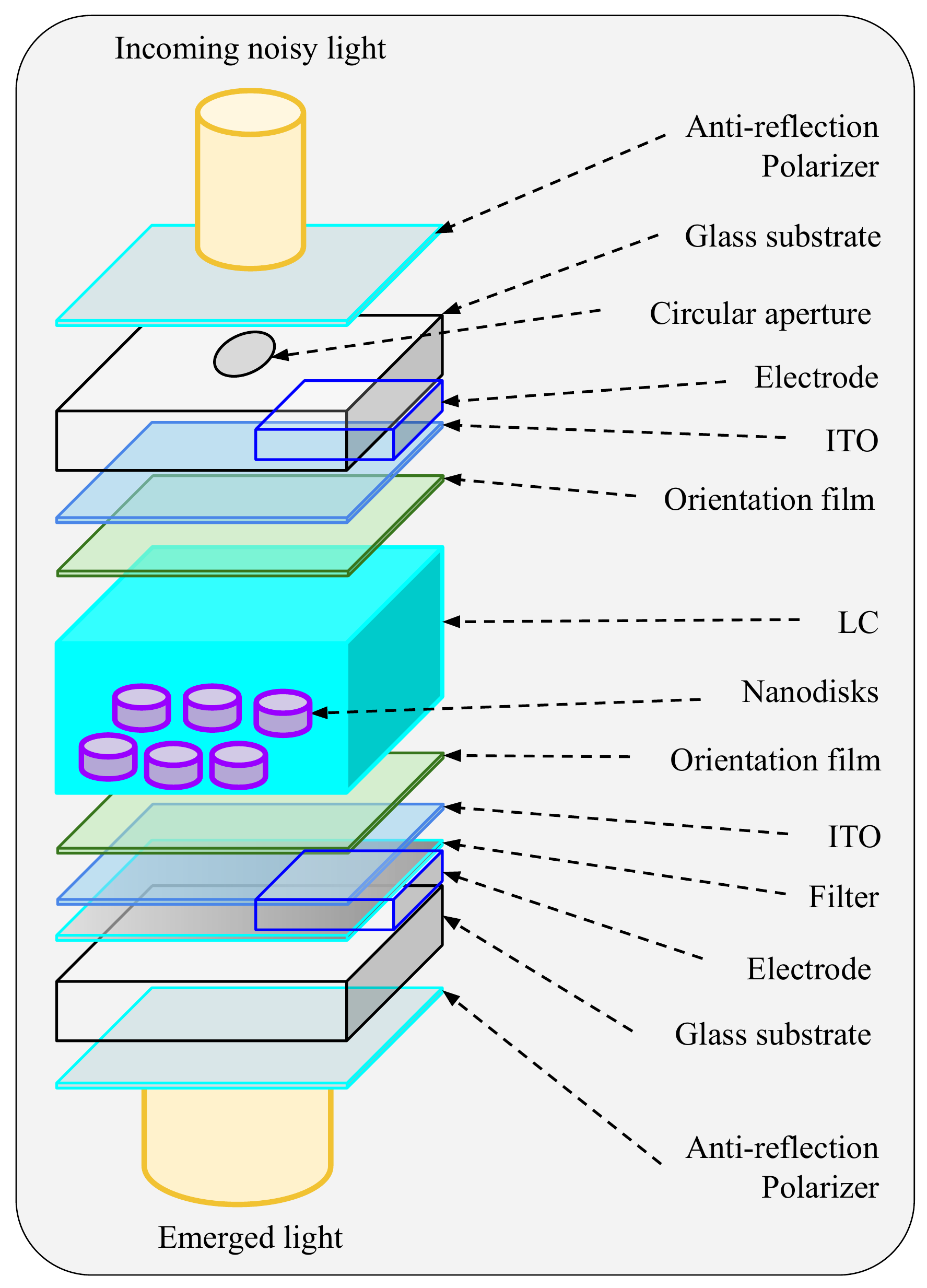}
		\caption{Structure and perspective view of an LC-based RIS for VLC receivers.}
		\label{fig:stimu_emission}
	\end{figure}

Figure \ref{fig:stimu_emission} depicts the structure and perspective view of an LC-based RIS element for VLC receivers, in which the construction is similar to most proposed designs \cite{alain, sasaki2015dynamic}. It shows a light conditioning unit using an LC cell and about twelve layers of thin materials located on both sides of the cell. They are the anti-reflection polarizer, glass substrate, electrode, the indium tin oxide (ITO) layer, and the orientation film. Of these materials, only the glass substrate can create a direction change of the light ray due to its thickness and refractive index, which is about 1.5. The anti-reflection polarizer filters the incoming light to let in only wavelengths carrying the transmitted information and deals with glare and skies. Its design depends on the LC's application. For example, indoor VLC receivers might not need skies dimming capabilities. In the case of multi-color transmissions, the polarizer may integrate color filtering elements. The glass substrates are in charge of generating the director\footnote{In a cell, LC’s molecules tend to point toward a preferred direction called the director.}, $ \hat{n} $, which defines the final light orientation when an external field is applied \cite{6626632}. Practically, the ITO is a thin conductive coating material realized with a fair trade-off between conductivity and transparency. Its primary role is to assist with heat generation and control for the LC cell. Finally, the orientation film guides light-rays through the preferred direction in the LC cell.
	
Dye-doped and polymer LCs mostly require an array of silicon nanodisks to minimize losses due to in-layer absorption and polarization based on the concept of silicon Huygens, which allows overlapping of electric and magnetic fields \cite{huyguens}. Note that the difference between dye-doped and polymer LCs resides mostly in the externally applied field. It may be lower for polymer LCs to achieve the same results as dye-doped LCs, in which the parameters are easily and smoothly tuned compared to polymer types. The material used must allow smooth tuning of its properties by the externally applied field and provide a high-quality factor, $ Q_{ris} $. The obtained device may behave like an amplifier, a resonator, a light filter, a refractive device, or a delay line. Varying the externally applied voltage through an efficient RIS algorithm undeniably affects most electro-optical and refractive characteristics of the sample, including the birefringence, $ \Delta n $, refractive index, $ n_{ris} $, electro-optic coefficient, $ r_{ris} $, permeability, and permittivity. As a result, the absorbance, $ A $, and transmittance, $ T_{ris} $, of the structure vary and explain why a distinct dye produces a particular result. Following the tuning process, the gain coefficient, $\Gamma_{\lambda}$, which characterizes the light-amplifier, also varies with the dye used, its concentration, the other electro-refractive parameters, and the externally applied voltage.
	\subsection{Effects of LC Parameters and Coefficients} 
In this subsection, we discuss the parameters that may vary under an externally applied electric field's influence, impacting the emerged light power.
	\begin{figure}
		\centering
		\includegraphics[width=0.46\textwidth]{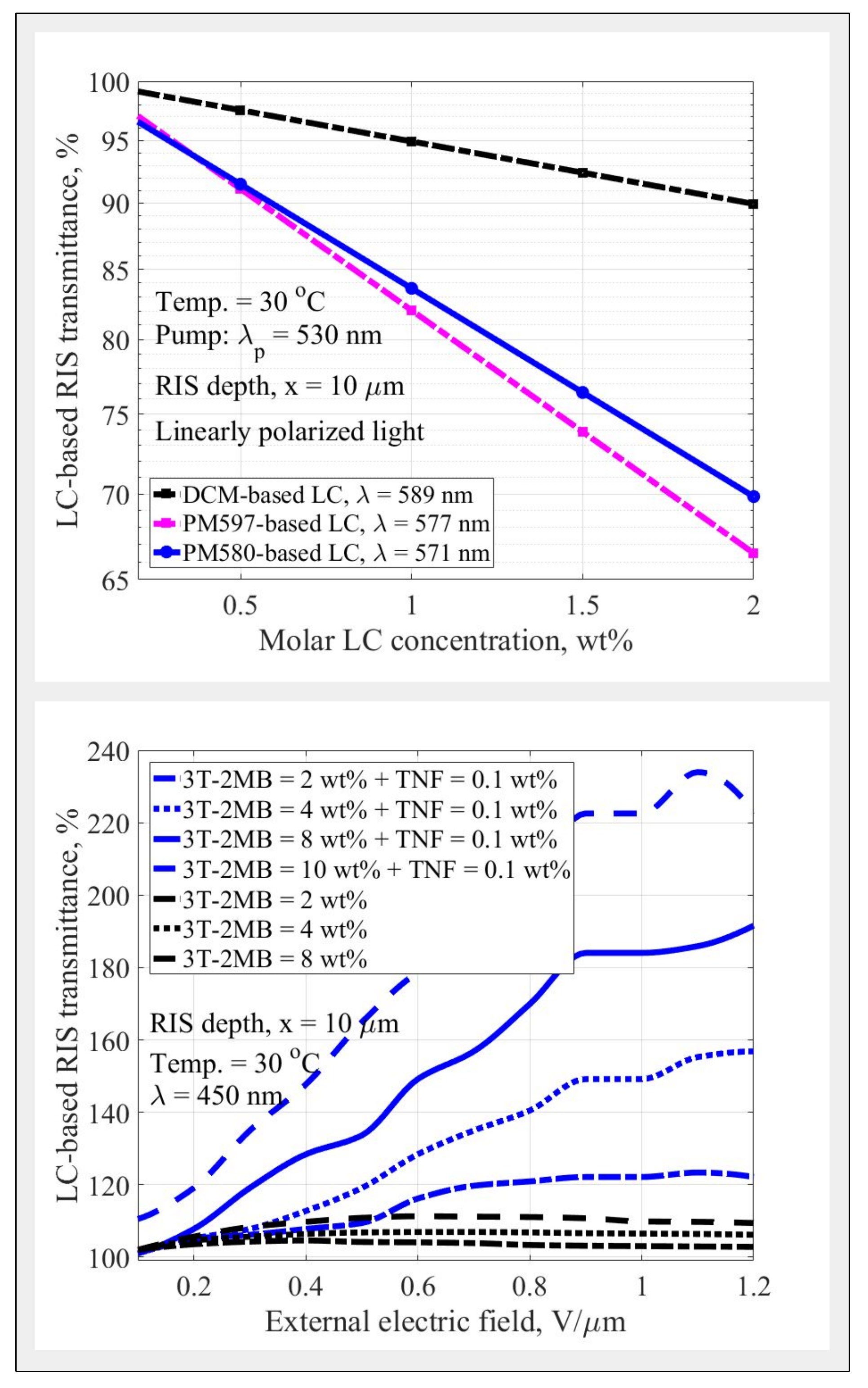}
		\caption{LC-based RIS transmittance in \%. Top: Transmittance versus LC molar concentration in weighted percentage (wt\%) \cite{mowatt2010comparison}. Bottom: Transmittance versus externally applied electric field, $ E_0 $ \cite{sasaki2014photorefractive}.}
		\label{fig:Tr_cons}
	\end{figure}
	\\
\textbf{LC concentration and dopant weighted percentage, wt\%}: The LC concentration affects the light attenuation as it goes through the RIS element. Consequently, it is part of the parameters that dictate light transmittance. Doping an LC with nano-particles creates a mixture which significantly impacts the electro-optic properties of the LC. For example, the permittivity of the mixture decreases as the nano-particle percentage increases. Even though the externally applied electric field is crucial, the dopant also changes the transition phase temperature, order parameters, and diffraction properties. Under natural conditions, with no external electric field, light transmittance through the RIS sample decreases as the mixture's concentration increases. 
	
\begin{figure}
		\centering
		\includegraphics[width=0.46\textwidth]{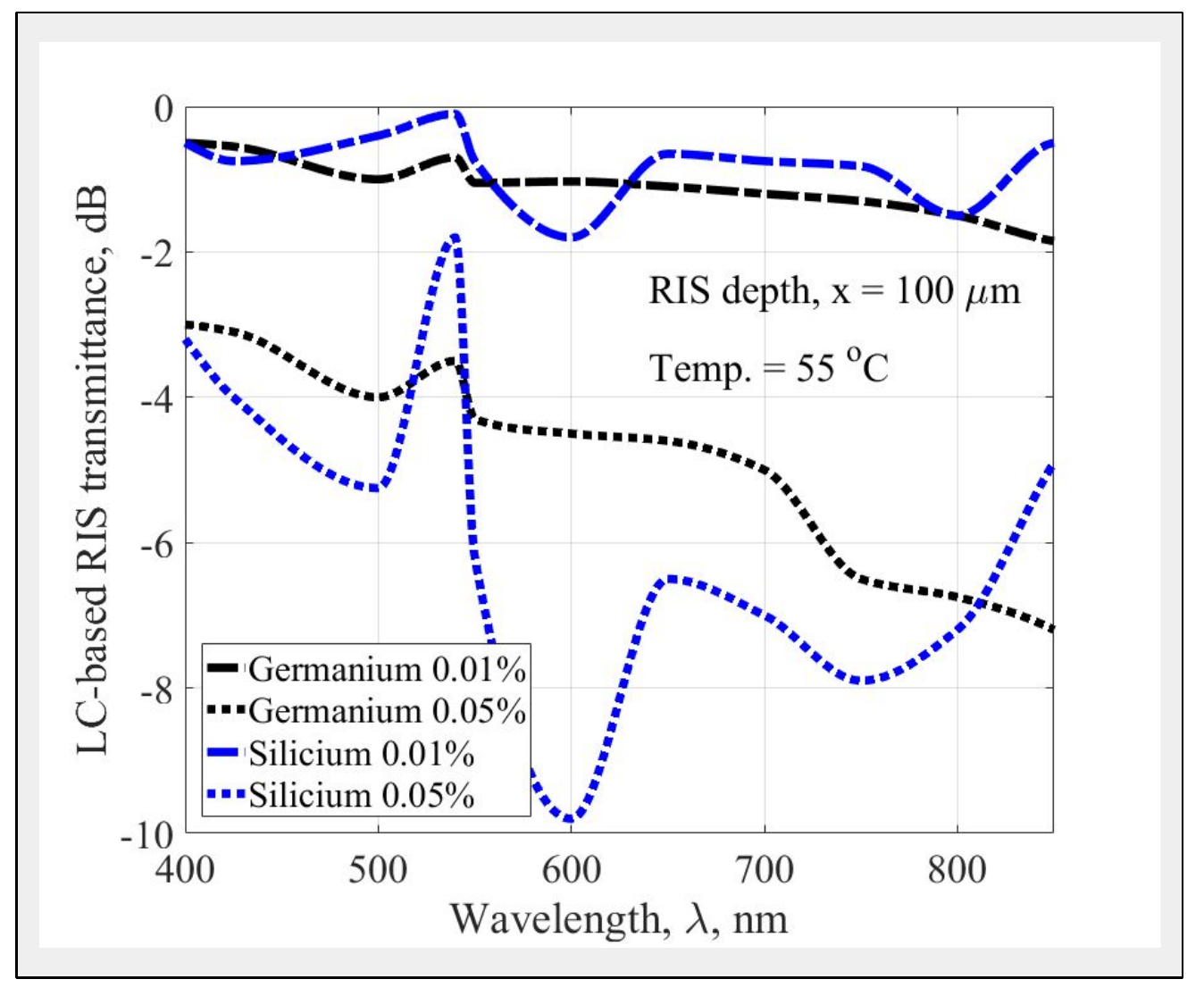}
		\caption{LC-based RIS transmittance of selected samples and dyes versus an externally applied electric vield, $ E_0 $, \cite{7042741}.}
		\label{fig:Gain_Coef}
	\end{figure}

Figures~\ref{fig:Tr_cons} and \ref{fig:Gain_Coef} show selected results obtained based on data retrieved from reports on experimental works presented in \cite{mowatt2010comparison}, \cite{sasaki2014photorefractive}, and \cite{7042741}, respectively. They highlight the influence of the RIS substance molar concentration, externally applied electric field, and wavelength on RIS transmittance. Figure~\ref{fig:Tr_cons} depicts the effects of the mixture concentration on light transmittance and amplification. The top-part shows the transmittance of three different dyes, including 4-Dicyanomethylene-2-methyl-6-p-dimethylaminostyryl-4H-pyran (DCM) at 589 nm, pyrromethene 580 (PM580) at 571 nm, and pyrromethene 597 (PM597) at 577 nm \cite{mowatt2010comparison}. It confirms that the transmittance decreases as the dye concentration increases; the incident light intensities (589, 571, and 577 nm) attenuate as the LC molar concentration increases. Applying an external source with variable frequency, polarity, or voltage may drastically reverse these slops. The bottom part of Fig.~\ref{fig:Tr_cons} shows the obtained transmittance percentage versus the externally applied electric field for different percentages of distinct LC mixtures. The LC cell contains a photoconductive chiral compound with the terthiophene, 3T-2MB LC, mixed with a 0.1 wt\% concentration of trinitrofluorenone (TNF). The obtained LC-based RIS thickness is $x$ = 10 $\mu$m. At 30$^oC$, it can significantly amplify a 450 nm incident light \cite{sasaki2014photorefractive}. The figure shows that increasing the externally applied electric field drastically increases the transmittance percentage, and thus the amplification gain.
\\
\textbf{Electro-optic coefficient, $ r_{ris} $}: An LC-based RIS structure is also seen as a voltage-controlled refractive index material because the application of an externally controlled voltage influences the speed of light traveling through it. It creates a perturbation that interacts with the light photons and results in properties change. The electro-optic coefficient, $ r_{ris} $, of the LC-based RIS quantifies the external voltage's effects on the element and helps define the amplification coefficient. Most used types of LC substances, such as nematic LC with a diffraction time enhancement, ferroelectric LC with short response time, and polymer dispersed LC, are susceptible to photo-refractive phenomenon. In such substances, the electro-optic coefficients are in general proportional to the emerged light power.     
	\\
\textbf{The refractive index, $ n_{ris} $}: It provides the RIS element with the capacity to guide the incoming rays, and is the main parameter offering LCs their wave-guiding potentials \cite{alain}. It includes two main components' contribution, namely, the ordinary refractive index, $ n_o $, and the extraordinary index, $ n_e $. Both components are temperature, wavelength, and voltage-dependent. When light propagates in parallel to the director, $ \hat{n} $, we observe an isotropic optical behavior of LC, and its birefringence nature occurs when the light ray is perpendicular to $ \hat{n}$, which makes $ n_{ris} $ a complex number \cite{6626632}. The birefringence, $ \Delta n $, combined with the LC-based RIS thickness, define the phase shift between incidents and observed rays. The refractive index is also one of the LC-based RIS amplification coefficients. For example, in the bottom part of Fig.~\ref{fig:Tr_cons}, the emerged amplified light is proportional to the cube of $ n_{ris} $ as explained in \cite{sasaki2015dynamic}.
	\\
\textbf{Amplification coefficient, $\Gamma_{\lambda}$}: In our application, the focus is on low-light amplification. Unlike other laser pumping techniques, we use an externally applied electric field to create a movement of particles in the LC substance. The electric field promotes the ground state population \cite{silfvast2004laser}. In practice, the obtained amplification relates to the photo-refractive theory, which stipulates a sign inversion of the attenuation coefficient, leading to an exponential emission \cite{sasaki2015dynamic}. The emerged light intensity, $I_e$, depends on the refraction order and index, wavelength, electro-optic coefficient, incidence angle, the externally applied electric field, and the RIS thickness, $x$. Among these parameters, the refractive index, $n_{ris}$, and the electro-optic coefficient, $r_{ris}$, are wavelength and external field-dependent. \cite{sasaki2015dynamic}. Figures~\ref{fig:Tr_cons} and \ref{fig:Gain_Coef}, which show the amplification gain in terms of the mixture concentration, externally applied electric field, and wavelength, are based on monochromatic incident light. In the case of a multi-color input wave, the emerged light color depends on mechanical stress, temperature, and externally applied electric field.
	\subsection{Effects of External Parameters} 
	\begin{table*}
		\begin{center}
			\caption{Emerged light power, $ P_e $ (mW), RIS gain (dB), and possible channel expansion, $ l $ (cm), for an LC-based RIS sample of thickness, $ x $ = 10 $ \mu $m, containing 4 to 8 wt\% of 3T-2MB and 0.01 wt\% of TNF, for an incident light power, $ P_i $ = 6 mW.}
			\label{tab:table1}
			\colorbox{lightgray!25}{
				\begin{tabular}{|l|c|c|c|c|c|c|c|c|c|c|c|c|}
					\hline
					\footnotesize{$ E_0 $ (V/$ \mu $m)} & 0.1 & 0.2 & 0.3 & 0.4 & 0.5 & 0.6 & 0.7 & 0.8 & 0.9 & 1.0 & 1.1 & 1.2\\
					\hline
					\multicolumn{13}{c}{4 wt\% 3T-2MB, 30$ ^oC $, $ x $ = 10 $ \mu $m, $ P_i $ = 6 mW} \\ \hline
					\footnotesize{$ P_e $ (mW)} & 6.0012 & 6.0027 & 6.0033 & 6.0037 & 6.0040 & 6.0040 & 6.0040 & 6.0040 & 6.0038 & 6.0038 & 6.0037 & 6.0036 \\
					$\Gamma_{\lambda}$ (dB) & 0.0009 & 0.0015 & 0.0018 & 0.0020 & 0.0018 & 0.0017 & 0.0017 & 0.0014 & 0.0013 & 0.0013 & 0.0013 & 0.0012 \\ \hline
					$ l $ (m) & 0.20 & 0.45 & 0.56 & 0.63 & 0.67 & 0.68 & 0.68 & 0.67 &  0.65 & 0.64 & 0.63 & 0.61 \\ \hline
					\multicolumn{13}{c}{8 wt\% 3T-2MB, 30$ ^oC $, $ x $ = 10 $ \mu $m, $ P_i $ = 6 mW} \\ \hline
					\footnotesize{$ P_e $ (nW)} & 6.0012 & 6.0033 & 6.0048 & 6.0056 & 6.0062 & 6.0064 & 6.0064 & 6.0063 & 6.0061 & 6.0056 & 6.0056 & 6.0054 \\
					$\Gamma_{\lambda}$ (dB) & 0.0009 & 0.0020 & 0.0024 & 0.0027 & 0.0029 & 0.0029 & 0.0029 & 0.0029 & 0.0028 & 0.0027 & 0.0027 & 0.0026 \\ \hline
					$ l $ (m) & 0.20 & 0.55 & 0.81 & 0.94 & 1,04 & 1.08 & 1.07 & 1.06 & 1.03 & 0.94 & 0.94 & 0.91\\ \hline
					\multicolumn{13}{c}{4 wt\% 3T-2MB + 0.01 wt\% TNF, 30$ ^oC $, $ x $ = 10 $ \mu $m, $ P_i $ = 6 mW} \\ \hline
					\footnotesize{$ P_e $ (mW)} & 6.0006 & 6.0030 & 6.0045 & 6.0072 & 6.0105 & 6.0150 & 6.0180 & 6.0204 & 6.0240 & 6.0240 & 6.0265 & 6.0271 \\
					$\Gamma_{\lambda}$ (dB) &  0.0004 & 0.0022 & 0.0026 & 0.0033 & 0.0039 & 0.0065 & 0.0078 & 0.0083 & 0.0087 & 0.0087 & 0.0091 & 0.0087 \\ \hline
					$ l $ (m) & 0.10 & 0.51 & 0.76 & 1.21 & 1.77 & 2.53 & 3.03 & 3.43 & 4.04 & 4.04 & 4.44 & 4.54\\ \hline
					\multicolumn{13}{c}{8 wt\% 3T-2MB + 0.01 wt\% TNF, 30$ ^oC $, $ x $ = 10 $ \mu $m, $ P_i $ = 6 mW} \\ \hline
					\footnotesize{$ P_e $ (mW)} & 6.0006 & 6.0045 & 6.0105 & 6.0150 & 6.0174 & 6.0240 & 6.0271 & 6.0319 & 6.0367 & 6.0367 & 6.0373 & 6.0391 \\
					$\Gamma_{\lambda}$ (dB) & 0.0004 & 0.0022 & 0.0033 & 0.0052 & 0.0076 & 0.0109 & 0.0130 & 0.0148 & 0.0174 & 0.0174 & 0.0191 & 0.0195 \\ \hline
					$ l $ (m) & 0.10 & 0.76 & 1.77 & 2.53 & 2.93 & 4.04 & 4.54 & 5.35 & 6.16 &  6.16 & 6.26 & 6.56\\ \hline
			\end{tabular}}
		\end{center}
	\end{table*} 	
\textbf{Externally applied Electric field, $ E_o $}: Note that an externally applied electric field, $ E_o $, controls the effects of all the above mentioned parameters. Since the LC-based RIS is either dielectric or ferroelectric high resistivity materials, it requires little energy to produce a significant effect. The dielectric nature of LC molecules allows these to align with the director, $ \hat{n} $, under the influence of $ E_o $. These molecules orient themselves with the electric field direction. This re-orientation requires the molecules to form permanent dipoles. However, even if molecules are not permanent dipoles, the externally applied electric field may re-arrange them to induce a permanent dipole characteristic. The response of an LC-based RIS to the externally applied electric field is observed on the RIS refractive index, electro-optic coefficient, permittivity, and permeability. A change in any of these leads to transmittance variation (see the bottom part of Fig.~\ref{fig:Tr_cons}). 
	\\ 
\textbf{Temperature}: The temperature has an indisputable effect on the LC-based RIS light amplification. As the LC cell temperature increases, the thermotropic LC goes through a series of phase transitions, which affects the RIS transmittance. In some temperature ranges, LC substances show both the properties of liquids and crystalline state. In practice, one may find that most of them are organic compounds with elongated molecules, in which the amplification coefficient decreases when the temperature increases \cite{sasaki2014photorefractive}.
	\\
\textbf{Wavelength, $ \lambda $}: Absorption coefficient and other LC parameters such as the refractive index are wavelength-dependent. In Fig.~\ref{fig:Gain_Coef}, we confirm that the response of an LC-based RIS varies with the wavelength, $ \lambda $ \cite{7042741}. The figure shows an LC-based RIS transmittance percentage for Germanium and Silicon based LC substances at 0.01 and 0.05 wt\% concentrations, respectively. The amplification pattern, which relates to the wavelength, follows the same trend for both concentration percentages of each material.
	\subsection{Figure of Merit} 
The quality factor, $ Q_{ris} $, defines the efficiency of light transition through an LC-based RIS structure. It dictates whether the receiver is in Bragg-diffraction or Raman-Nath regimes. Note that the Raman-Nath regime allows several diffraction orders, whereas the Bragg-diffraction regime only allows one order of diffraction. To avoid more reflections within the LC-based RIS structure, we impose a single diffraction order (Bragg-diffraction regime), with $ Q_{ris} > 1 $. Most designs use a value of $ Q_{ris} $ greater than 10 to ensure a Bragg-diffraction regime. 
	\subsection{Amplification Gain and Modes of Operation}
Based on the transmittance showed in the bottom part of Fig.~\ref{fig:Tr_cons}, we calculate the emerged light power, $ P_e $, for an input signal, $ P_i $ = 6 mW. We evaluate the emerged light power, $ P_e $, and distance, $ l $, at which the emerged signal attenuates to the value of $ P_i $. Table~\ref{tab:table1} shows these results against different values of the externally electric field. The table also gives the RIS gain for various values of the externally applied electric field, $ E_o $, and highlights the VLC transmission range expansion. We assume that a monochromatic incident light of power, $ P_i $ = 6 mW, penetrates the RIS structure. It is considered that $ \zeta_{air} $ = 0.43 dB, and $ l $ is determined using the Beer-Lambert law. We also consider that $ P_i $ = 6 mW is the minimum light power the PD can detect. We present results for the selected mixtures of LC with 4 and 8 wt\% 3T-2MB with and without TNF. An RIS thickness $ x $ = 10 $ \mu $m, and an LC temperature of 30$ ^o $C are assumed. Results undeniably show, as Fig.~\ref{fig:Tr_cons} confirms, that adding a small TNF percentage to the LC mixture with 3T-2MB improves the gain coefficient. For systems without TNF, we obtain an amplification with gain between 0.0009 and 0.0020 dB for 4 wt\% 3T-2MB, and between 0.0009 and 0.0029 dB for 8 wt\% 3T-2MB, which leads to a possible transmission range expansion of 0.20 to 0.68 m and 0.20 to 1.08 m, respectively. Similarly, we obtain an amplification gain between 0.0004 and 0.0091 dB for 4 wt\% 3T-2MB plus 0.1 wt\% TNF and between 0.0004 and 0.0195 dB for 8 wt\% 3T-2MB plus 0.1 wt\% TNF, which leads to 0.10 to 4.54 m and 0.10 to 6.56 m expansion range, respectively.
	
The obtained increment of the VLC transmission range can be implemented in two scenarios; enhanced VLC detected signal and direct VLC signal amplification.
	\\  
\textbf{Enhanced VLC detected signal}: In this mode of operation, the PD and LC-based RIS are in the same package; the PD is located at the bottom of the LC cell. The LC-based RIS structure should handle amplification in such a way to prevent the PD from working in the saturation region, and thus avoid higher distortion of the received signal. Under this scenario, the LC-based RIS improves the VLC range by a few meters. Mixtures of 4 and 8 wt\% 3T-2MB may adapt well in this situation as shown in Table~\ref{tab:table1}.
	\\
\textbf{Direct VLC signal amplification}: In this case, the PD and VLC receiver are situated at a distance, $ l \gg x$, from the RIS structure, which acts as a relay. The emerged light travels the distance $ l $ to reach the PD through the air with an attenuation coefficient, $ \zeta_{air} $. Without the use of the LC-based RIS, the light from the VLC transmitter will not reach the PD. We regard the system as two subsystems; one before and one after the RIS structure. Note that energy can be harvested from the channel to power the LC-based RIS module as it requires a low-power source to operate.
	\section{Future Research Directions and Conclusion}
Using a RIS element in VLC receivers to replace the traditional concentrator based on convex lens is novel and requires further investigation. It is crucial to consider analyzing VLC systems with a RIS element for specific VLC applications. For example, VLC applications requiring point-to-point protocols and intelligent transportation systems. Practical implementation of these new VLC systems should be analyzed and explored to enable more applications of LC-based RIS in VLC receivers. 
	
Underwater is one of the communications environments where the attenuation coefficient and sliding effects are higher for most waves in the electromagnetic spectrum. It represents one of the transmission environments where LC-based RIS would find irrefutable applications and therefore requires further investigations. The selection of adequate LC substances, polymer, or dye types also requires further analysis for these VLC applications. 
	
	\vskip 0.04in
	
This article focuses on using LC-based RIS in VLC systems to replace the traditional concentrator based on convex lenses, with the aim of improving the VLC transmission range. We discussed the effect of selected LC mixture properties on incident monochromatic light rays. We also elaborated on the externally applied electric field's effects on the LC mixture's electro-refractive coefficients. We showed that using LC-based RIS in VLC systems can improves the transmission range by up to 6.56 m, depending on the type of LC, mixture and concentration, and the externally applied electric field. Finally, we outlined research opportunities towards the use of LC-based RIS in VLC systems.

	\ifCLASSOPTIONcaptionsoff
	\newpage
	\fi
	
	\bibliographystyle{IEEEtran}
	\bibliography{meta_optics}
	\vskip -1\baselineskip plus -1fil
	\begin{IEEEbiographynophoto}{Alain R. Ndjiongue}[S'14, M'18, SM'20]
		received the M.Eng. and D. Eng degrees in electrical and electronic engineering from the University of Johannesburg, South Africa, in 2013 and 2017, respectively. He served as a PDRF and senior lecturer at the same university from 2017 to 2020, and is currently a senior researcher at Memorial University of Newfoundland, Canada. His research interests are in digital communications, including power line and optical communications. He is an active reviewer for several high impact IEEE ComSoc journals and conferences. 
		He was a 2017 and 2018 exemplary reviewer for IEEE Communication Letters, Optical Society of America, and a 2019 top 1\% peer reviewer in the Essential Science Indicators Research.
	\end{IEEEbiographynophoto}
	\vskip -2.4\baselineskip plus -1fil
	\begin{IEEEbiographynophoto}{Telex. M. N. Ngatched} [M'05, SM'17]
		is an associate professor at the Grenfell Campus at Memorial University, Canada. His research interests include 5G enabling technologies, visible light and power-line communications, optical communications for OTN, and underwater communications. He is an associate editor of the IEEE Open Journal of the Communications Society and the Managing Editor of IEEE Communications Magazine. He was the publication chair of IEEE CWIT 2015, an associate editor for IEEE Communications Letters from 2015 to 2019, as well as Technical Program Committee member and session chair for many prominent IEEE conferences in his area of expertise.
	\end{IEEEbiographynophoto}
	\vskip -2.4\baselineskip plus -1fil
	\begin{IEEEbiographynophoto}{Octavia A. Dobre} [M’05, SM’07, F’20]
		is a professor and Research Chair at Memorial University, Canada. She was a visiting professor at Massachusetts Institute of Technology, as well as a Royal Society and a Fulbright Scholar. Her research interests include technologies for 5G and beyond, as well as optical and underwater communications. She has published over 300 referred papers in these areas. She serves as the Editor-in-Chief (EiC) of the IEEE Open Journal of the Communications Society. She was the Editor-in-Chief of IEEE Communications Letters, a senior editor and an editor with prestigious journals, as well as General Chair and Technical Co-Chair of flagship conferences in her area of expertise. She is a Distinguished Lecturer of the IEEE Communications
		Society and a fellow of the Engineering Institute of Canada.
	\end{IEEEbiographynophoto}
\vskip -2.4\baselineskip plus -1fil
\begin{IEEEbiographynophoto}{Harald Haas} [S'98, AM'00, M'03, SM'16, F'17] received the Ph.D. degree from the University of Edinburgh in 2001. He currently holds the Chair of Mobile Communications at the University of Strathclyde, and is the initiator, co-founder and Chief Scientific Officer of pureLiFi Ltd as well as the Director of the LiFi Research and Development Center at the University of Strathclyde. His main research interests are in optical wireless communications, hybrid optical wireless and RF communications, spatial modulation, and interference coordination in wireless networks.
\end{IEEEbiographynophoto}		
\end{document}